\documentclass[prb,twocolumn,preprintnumbers,amsmath,amssymb, showkeys]{revtex4}
\usepackage{graphicx}
\usepackage{dcolumn}
\usepackage{bm}
\usepackage{titlesec}           
\usepackage{titletoc}

\begin{document}
\title{1-D Simulation of the Electron Density Distribution in a Novel Nonvolatile Resistive Random Access Memory Device}
\author{Ren\'e Meyer and~Hermann Kohlstedt}
\address{Institut f\"ur Festk\"orperforschung (IFF) and CNI, the Center of Nanoelectronic Systems for Information Technology Forschungszentrum J\"ulich, 52425 J\"ulich,
Germany}

\begin{abstract}
The operation of a novel nonvolatile memory device based on a conductive ferroelectric/non-ferroelectric thin film
multilayer stack is simulated numerically. The simulation involves the self-consistent steady state solution of
Poisson's equation and the transport equation for electrons assuming a Drift-Diffusion transport mechanism. Special
emphasis is put on the screening of the spontaneous polarization by conduction electrons as a function of the applied
voltage. Depending on the orientation of the polarization in the ferroelectric layer, a high and a low resistive state
are found giving rise to a hysteretic ${I}$-${V}$ characteristic. The ${R_{\rm high}}$ to ${R_{\rm low}}$ ratio ranging
from $>$ 50\% to several orders of magnitude is calculated as a function of the dopant content.
\end{abstract}


\maketitle

\section{Introduction}
Nonvolatile memories based on ferroelectric capacitor structures (FRAM) are considered as a potential candidate to
replace the current Flash memory generation \cite{FRAM1, FRAM2, Scott}. However, the ongoing trend to miniaturization
may level the technological efforts in bringing FRAM into production, since charge based devices require a minimum area
for a save information readout. Various concepts of {\it resistive} memories (RRAM) e.g. based on magnetic tunnel
junctions (MRAM) \cite{MRAM}, phase change materials (PCM, Ovonics) \cite {PCM, Ovonics}, ferroelectric tunnel
junctions \cite{Kohlstedt}, ferroelectric diodes \cite{Watanabe, Blom, Sluis} or charge trapping/ detrapping effects
\cite{Bednorz}, which do not show this disadvantage, are currently under development. The key issue of all approaches
is to overcome the limitations of today's Flash memories concerning writing speed, cycling endurance and large
programming voltages.

This contribution reports on numerical simulation studies of a memory device consisting of a planar conductive
ferroelectric/ non-ferroelectric multilayer stack, which combines the ferroelectric information storage principle with
a low resistive information readout. In the following, it will be referred to as \underline{F}erro
\underline{R}esistive \underline{R}andom \underline{A}ccess \underline{M}emory, FRRAM. The storage mechanism is based
on the establishment of potential well or a potential barrier for electrons, which can be controlled by polarization
reversal in the ferroelectric layer. This causes an enrichment or depletion of the majority charge carrier electrons at
the ferroelectric/non-ferroelectric interface, whereby the structure is electrically characterized by a high and a low
resistive state. In contrast to a ferroelectric diode concept, the proposed device consists of {\it two highly
conductive} thin film layers sandwiched between two {\it low ohmic} contacts.

\section{Model}
Fig.\,1(a) shows the capacitor-like memory device structure consisting of a ferroelectric and a non-ferroelectric layer
in contact with metal electrodes. To elucidate the operation principle of the device, we first regard the ferroelectric
and the non-ferroelectric layer to be ideal insulators. In a refined model approach, a high concentration of conduction
electrons will be assumed to be present in both layers.

\begin{figure}[b]
\centering
\includegraphics[width=2.2in]{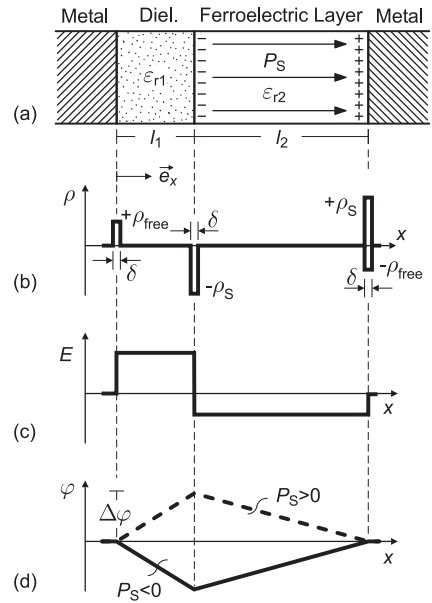}
\caption{Simplified operation principle of the proposed memory device consisting of (a) an ideal ferroelectric/
insulator stack. (b) space charge distribution $\rho$ in the multilayer, (c) the resulting inner electric field $E$ and
(d) inner potential $\varphi$ for different orientations of the spontaneous polarization. Depending on the orientation
of ${P_{\rm S}}$ in the ferroelectric layer, a {\it potential barrier} or a {\it potential well} for electrons is
formed.} \label{model_simplified}
\end{figure}

\subsection{Ideal Ferroelectric/ Insulator Multilayer}
In the absence of mobile charge carrier in the multilayer, the space charge distribution $\rho(x)$ for short circuit
conditions is given by
\begin{equation}
\rho (x) = \left\{ \begin{array}{cl} \rho_{\rm free} & \qquad x=0 \\
                                        \rho_{\rm S}              & \qquad x=l_1 \\
                                        \rho_{\rm free}-\rho_{\rm S}   & \qquad x=l_1+l_2 \\
                                        0 & \qquad else
\end{array}
\right.,
\end{equation}
whereby $\rho_{\rm free}$ denotes the influenced charge on the metal electrode and $\rho_{\rm S}$ is the spontaneous
polarization charge of the ferroelectric as shown in Fig.\,1(b). The relationship between space charge and the inner
electric field $E$ or the electric potential $\varphi$ are given by Maxwell's first equation and the material equation
\begin{equation}
\frac{dE}{dx} = \frac{\rho}{\varepsilon}
\end{equation}
and by Poisson's equation
\begin{equation}
\frac{d^2 \varphi}{dx^2} = - \frac{\rho}{\varepsilon},
\end{equation}
where $\varepsilon$ is the permittivity of the respective material. From the space charge distribution, the inner
electric field and the inner potential are estimated (see Fig.\,1(c)-(d)). It is found that the shape of the inner
potential is strongly affected by the orientation of the polarization. For ${{\vec P}_{\rm S} \uparrow \uparrow {\vec
e}_x}$ we predict the formation of a {\it potential barrier}, whereas a {\it potential well} is formed for ${{\vec
P}_{\rm S} \uparrow \downarrow {\vec e}_x}$. Here, ${{\vec e}_x}$ denotes the normal vector in $x$-direction. Both
states can be addressed by application of an external voltage of respective polarity, so that the voltage drop across
the ferroeletric layer exceeds the coercive voltage of the ferroelectric material. The present state of the potential
barrier can easily be sensed by measuring the conductivity when conduction electrons e.g. by doping are introduced into
the structure. The latter is expected to cause a screening of the polarization charge. Electrons will then be attracted
by the positive polarization charge and repulsed from the negative polarization charge. The question, whether or not
the shape of the potential estimated for an insulation system (see Fig.\,1(d)) will significantly change, if electrons
and donor atoms are present, will be tackled in the next subsection. We will focus on the redistribution of electrons
in an n-type semiconductor multilayer structure as a function of the orientation of the polarization parallel and
anti-parallel to an electronic current, which is driven by an external voltage.

\begin{figure}
\centering
\includegraphics[width=3.5in]{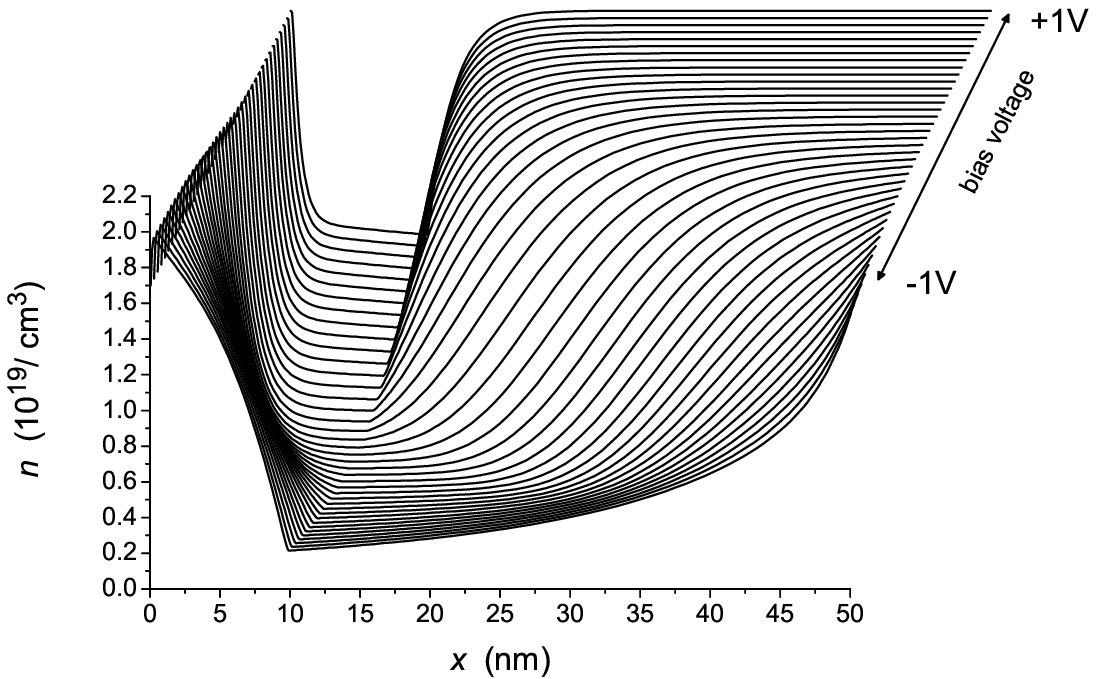}
\caption{Electron density distribution as a function of the applied voltage: I, case of enrichment (potential well).}
\label{n_enrichment}

\includegraphics[width=3.7in]{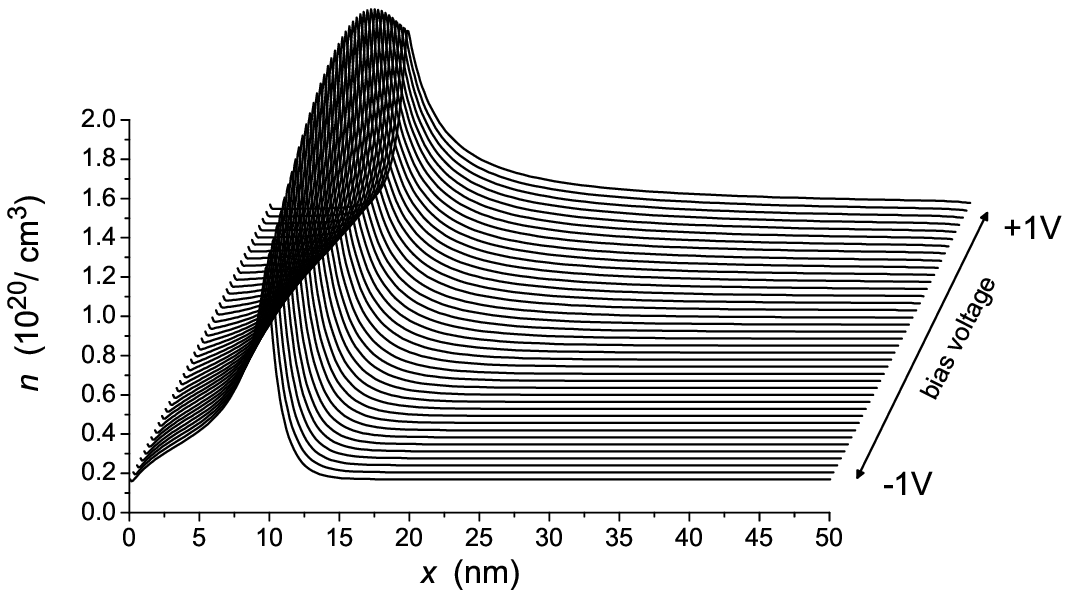}
\caption{Electron density distribution as a function of the applied voltage: II, case of depletion (potential
barrier).} \label{n_depletion}
\end{figure}
\subsection{Conducting Ferroelectric/ Semiconductor Multilayer}
If conduction electrons are present in the structure due to shallow donor impurities, Eq.\,(1) has to be replaced with
\begin{equation}
\rho (x) = \left\{ \begin{array}{cl} \rho_{\rm free} & \qquad x=0 \\
                                        \rho_{\rm S}              & \qquad x=l_1 \\
                                        \rho_{\rm free}-\rho_{\rm S}   & \qquad x=l_1+l_2 \\
                                        e_0(n-N_{\rm D}^+) & \qquad else
\end{array}
\right..
\end{equation}
Here, $e_0$ is the elementary charge, $n$ denotes the concentration of conduction electrons and $N_{\rm D}^+$ is the
concentration of ionized donor atoms. Following a Drift-Diffusion approach, the redistribution of electrons under the
inner electrical field and a concentration gradient is given by \cite{Fridkin}
\begin{equation}
j_n = - D_n \frac{dn}{dx} - \mu_n n E,
\end{equation}
where $D_n$ is the diffusion coefficient of the electrons, $\mu_n$ the electron mobility and $j_n$ the current density.
Under {\it steady state} conditions, the transport equation simplifies to
\begin{equation}
\frac{dn}{dt} = \frac{dj_n}{dx} \equiv 0.
\end{equation}
Low ohmic contact properties of the electrode interfaces were approximated by neutral contacts. Then, the boundary
conditions for electrons are given by
\begin{equation}
n(x=0) = n(x=l_1+l_2) \equiv N_{\rm D}^+.
\end{equation}

The potential difference between the system boundaries is determined by the applied external voltage. For the sake of
simplicity, a homojunction between the ferroelectric and the semiconducting layer and identical electrodes on both
sides are assumed
\begin{equation}
\varphi(x=l_1+l_2)-\varphi(x=0)=V_{\rm extern}.
\end{equation}
Under conservation of charge in the system, the redistribution of electrons can be calculated as a function of the
external voltage and the orientation of the spontaneous polarization.

\section{Results}
Numerical simulation techniques are applied to determine the self-consistent solution of Eqs.\,(2)-(8). The electron
profile for ${\rm -1V< \it {V}_{\rm extern}< \rm {1V}}$ for charge carrier enrichment is shown in Fig.\,2. The case of
depletion of electrons as a function of the applied voltage is illustrated in Fig.\,3. Both ferroelectric and
semiconducting layer have a donor concentration of $N_{\rm D}=10^{19}{\rm cm^{-3}}$ and the spontaneous polarization is
set to $|P_{\rm S}|={\rm 10\mu C/cm^2}$. Note: A switching of the ferroelectric polarization has not yet been
considered. For $P_{\rm S}>0$, a strong enrichment of electrons leads to a highly conductive state. In case of $P_{\rm
S}<0$, the depletion of electrons at the ferroelectric/ semiconductor interface correlates with a poorly conductive
state. As shown in the simulation, even if a large number of electrons is present in the ferroelectric layer, the
polarization charge is not perfectly screened, but still significantly affects the concentration of electrons in the
system. Although the inner potential reveals a certain distortion of the triangular shape as shown in Fig.\,1d, which
is caused by a band bending, the formation of a potential barrier or a potential well can still be observed.

\begin{figure}
\centering
\includegraphics[width=2.8in]{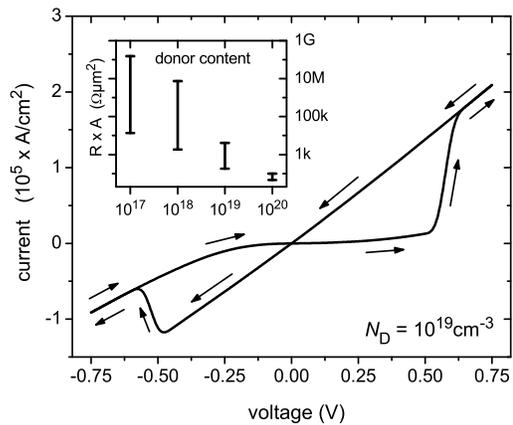}
\caption{Estimated $I$-$V$ characteristic for a donor concentration of $N_{\rm D}^+=10^{19}{\rm cm}^{-3}$. Inset:
Scalability of the resisitivity and $R_{\rm high}$ versus $R_{\rm low}$ ratio as a function of the dopant
concentration.} \label{I_V}
\end{figure}

An estimation of the $I$-$V$ curve from the electron distribution under consideration of a switching of the
ferroelectric polarization at an external voltage exceeding 0.5V and an electron mobility of $\mu_{\rm n }=1 {\rm
cm^2/Vs}$ is illustrated in Fig.\,4. A hysteretic resistance curve is observed, whereby the high or the low resistive
state can be read out by a voltage below the threshold voltage. The inset illustrates the scalability of the
resisitivity under variation of the donor content. Further adjustment can be achieved e.g. by variation of the layer
thicknesses, the value of the spontaneous polarization (different material systems) or the contact properties of the
electrode interfaces. Even materials with a small spontaneous polarization, which can not be used in FRAM applications,
may be suitable for the proposed memory concept. We regard complex oxides (Perovskites) \cite{Scott}, II-VI mixed
crystals (ZnCdS) \cite{Sluis}, polymers (PVDF-TFE) \cite{PVDF}, but also liquid crystals as possible candidates for the
realization of the FRRAM device.

\section{Conclusion}
We have demonstrated by numerical simulation studies the resistive storage properties of a novel memory device built
from a conducting ferroelectric semiconductor bilayer structure. A high and a low resistive state of the memory can be
addressed by switching the polarization in the ferroelectric layer. Special emphasis was put on the interaction between
conduction electrons and the ferroelectric leading to a redistribution of the majority charge carrier and a partial
screening of spontaneous polarization. An estimation of the $I$-$V$ characteristic as well as the influence of the
dopant concentration shows a wide scalability of the resistance. This illustrates the potential of the proposed device
as an alternative concept for future high integrated non-volatile random access memories ranging from Si-based
technology to polymer applications.

\section*{Acknowledgment}
The authors would like to thank N. A. Pertsev, R. Waser, ${\mbox{K. Szot}}$ and R. Oligschlaeger (the flying dutchman)
for fruitful discussions. Special thanks to A. R\"udiger for carefully reading the manuscript.

\end{document}